%% file: SMP-12-007_temp.tex
\pdfoutput=1

\documentclass[11pt,twoside,a4paper,cmspaper,final,collab]{cms-tdr}
\def\svnVersion{176987}\def\cmsCernNo{2012-336}\def\cmsCernDate{\today}\def\cmsMessage{Submitted to the Journal of High Energy Physics}

\begin{document}\cmsNoteHeader{SMP-12-007}

\hyphenation{had-ron-i-za-tion}
\hyphenation{cal-or-i-me-ter}
\hyphenation{de-vices}

\RCS$Revision: 159544 $
\RCS$HeadURL: svn+ssh://svn.cern.ch/reps/tdr2/papers/SMP-12-007/trunk/SMP-12-007.tex $
\RCS$Id: SMP-12-007.tex 159544 2012-11-29 14:23:51Z asavin $
\newlength\cmsFigWidth
\ifthenelse{\boolean{cms@external}}{\setlength\cmsFigWidth{0.85\columnwidth}}{\setlength\cmsFigWidth{0.4\textwidth}}
\ifthenelse{\boolean{cms@external}}{\providecommand{\cmsLeft}{top}}{\providecommand{\cmsLeft}{left}}
\ifthenelse{\boolean{cms@external}}{\providecommand{\cmsRight}{bottom}}{\providecommand{\cmsRight}{right}}
\newcommand{\tauh}{\ensuremath{\Pgt_\mathrm{h}}}
\newcommand{\taumu}{\ensuremath{\Pgt_{\Pgm}}}
\newcommand{\taue}{\ensuremath{\Pgt_{\Pe}}}
\newcommand{\taus}{\ensuremath{2\ell 2\Pgt}}
\newcommand{\PZ}{\cPZ}
\hyphenation{ATGCs}
\newcommand{\fourl}{\ensuremath{\ell^{+}_\mathrm{i} \ell^{-}_\mathrm{i} \ell^{+}_\mathrm{j} \ell^{-}_\mathrm{j}}}
\newcommand{\ZZfourl}{\ensuremath{{\PZ\PZ} \rightarrow 2\ell 2\ell}}
\newcommand{\ZZtwoltwotau}{\ensuremath{{\PZ\PZ} \rightarrow 2l2\tau}}
\providecommand{\mH}{\ensuremath{m_{\PH}\xspace}}
\renewcommand{\lumi}{\ensuremath{\,\text{(lumi.)}}\xspace}
\cmsNoteHeader{SMP-12-007} 
\title{Measurement of the $\cPZ\cPZ$ production cross section and search for anomalous couplings in  $2\ell2\ell'$ final states in pp collisions at $\sqrt{s} = 7\TeV$ }

\date{\today}

\abstract{
A measurement is presented of the $\cPZ\cPZ$ production cross section in the $\cPZ\cPZ \to 2\ell 2\ell'$ decay mode with
$\ell = \Pe, \Pgm$ and $\ell' = \Pe, \Pgm, \Pgt$
in proton-proton collisions at $\sqrt{s} = 7\TeV$ with the CMS experiment
at the LHC.
Results are based on data corresponding to an integrated luminosity of $5.0\fbinv$.
The measured cross section
$\sigma (\Pp\Pp \to \cPZ\cPZ) =  6.24 \,^{+0.86}_{-0.80} \stat \,^{+0.41}_{-0.32} \syst \pm 0.14 \lumi \unit{pb}$
is consistent with the standard model predictions.
The following limits on $\cPZ \cPZ \cPZ$ and $\cPZ \cPZ \gamma$ anomalous trilinear gauge
couplings
are set at 95\% confidence level:
$-0.011<f_4^\cPZ<0.012$, $-0.012<f_5^\cPZ<0.012$, $-0.013<f_4^{\gamma}<0.015$, and
$-0.014<f_5^{\gamma}<0.014$.
}

\hypersetup{%
pdfauthor={CMS Collaboration},%
pdftitle=
{Measurement of the ZZ production cross section and search for anomalous couplings in 2l2l' final states in pp collisions at sqrt(s)=7 TeV},%
pdfsubject={CMS},%
pdfkeywords={CMS, physics, software, computing}}

\maketitle 

The study of diboson production in proton-proton collisions provides an important
test of the standard model (SM). Many extensions of the SM predict new scalar,
vector, or spin-2 particles that decay into a pair of $\PW$ or $\cPZ$ bosons.
In addition, these
final states are sensitive to the self-interactions among the gauge bosons via trilinear
gauge couplings (TGCs). These couplings are the direct consequence of the non-Abelian
$SU(2) \times U(1)$ gauge symmetry of the SM and are a necessary
ingredient to construct renormalizable theories.
The values of these couplings are fully determined in the SM by the gauge structure
of the Lagrangian. Therefore, any deviation of the observed coupling strength
from the SM prediction would indicate the presence of new physics.
This deviation would be manifested as a change in the production cross section, especially
for energetic heavy gauge bosons.

In the SM, $\cPZ\cPZ$ production proceeds via the $t$- and $u$-channel
$\cPq\cPaq$ scattering
diagrams, and via gluon-gluon fusion.
The presence of anomalous neutral trilinear couplings (ATGCs)
would lead to a sizable enhancement of
$\cPZ\cPZ$ final states via $s$-channel $\cPq\cPaq$ scattering.
A model featuring such couplings can be constructed by means of
an effective Lagrangian~\cite{Hagiwara}.
In this parametrization, two $\cPZ\cPZ\cPZ$ couplings and two $\cPZ\cPZ\gamma$ couplings are
allowed by electromagnetic gauge invariance and Lorentz invariance for on-shell
Z bosons. The couplings are parametrized by two CP-violating ($f_4^{\cPZ, \gamma}$) and
two CP-conserving ($f_5^{\cPZ, \gamma}$) complex parameters, which are zero in the SM.

Measurements of the $\cPZ\cPZ$ cross section were previously performed at the Tevatron~\cite{CDFZZxs,D0ZZxs}
and the Large Hadron Collider (LHC)~\cite{HZZ4lPRL,ATLAS}.
A first measurement of the $\cPZ\cPZ$ cross section at center-of-mass energy
$\sqrt{s} = 7\TeV$  by the
Compact Muon Solenoid (CMS) Collaboration in the
decay mode $\ZZfourl$, where
$\ell$ is either $\Pe$ or $\Pgm$,
is presented in Ref.~\cite{HZZ4lPRL}. The measured cross section
$\sigma (\Pp\Pp \to \cPZ\cPZ) {\cal B} (\ZZfourl) =  28.1 \,^{+4.6}_{-4.0} \stat \, \pm 1.2 \syst \pm 1.3 \lumi \unit{fb}$ agrees well with the SM prediction of $27.9 \, \pm 1.9 \unit{fb}$.
In this Letter, we present an extended measurement
of the $\PZ\PZ$ production cross section based on the decay
mode $2\ell 2\ell'$, where
$\ell'$ is $\Pe$, $\Pgm$, or $\Pgt$.
If a $\Pgt$ is present in the final state, one $\PZ$ is required to decay
into
$\Pep\Pem$
or $\Pgmp\Pgmm$,
and the second
$\PZ$ into $\Pgt^+\Pgt^-$ in four possible final states: $\tauh \tauh$, $\taue \tauh$, $\taumu \tauh$, and
$\taue \taumu$,
where $\tauh$ represents a $\Pgt$  decaying
hadronically, while $\taue$ and $\taumu$ indicate taus decaying into an electron and a muon, respectively.
The presence of four leptons in the final state provides a clean signature with only a small
contribution from background processes.
The background sources include reducible contributions from $\cPZ\cPqb\cPaqb$ and $\ttbar$ processes,
where the final states contain two isolated leptons and two $\cPqb$ jets with secondary leptons,
and from $\cPZ$+jets and $\PZ\PW$+jets processes where the jets are misidentified
as leptons.

In this Letter, we also present
a search for the neutral $\cPZ\cPZ\cPZ$ and $\cPZ\cPZ\gamma$ ATGCs.
Previous studies on neutral $\cPZ\cPZ\cPZ$ and $\cPZ\cPZ\gamma$ ATGCs were performed at LEP2~\cite{ALEPH, DELPHI, LEPWG, OPAL, L3},
the Tevatron~\cite{D0}, and the LHC~\cite{ATLAS}.
The most restrictive limits were
set in Ref.~\cite{ATLAS},
$-0.07 < f_{4,5}^\cPZ< 0.07$ and
$-0.08 < f_{4,5}^\gamma < 0.08$,
with a data set corresponding to
an integrated luminosity of $1\fbinv$ of pp collisions at $\sqrt{s} = 7\TeV$.

The measurements presented here are based on
data collected in 2010 and 2011 with the CMS experiment at the LHC
at $\sqrt{s} = 7\TeV$, corresponding to an
integrated luminosity of $5.0 \pm 0.1\fbinv$.
A set of Monte Carlo (MC) event samples is used
to simulate signal and
background events.
The $\cPZ\cPZ$ production via $\cPq\cPaq$ is generated at next-to-leading order (NLO) with
\POWHEG2.0~\cite{Alioli:2008gx,Nason:2004rx,Frixione:2007vw}, while
other diboson processes (\PW\PW, \PW\cPZ, $\cPZ\gamma$) are generated with \PYTHIA6.424 and
\MADGRAPH~\cite{Madgraph5}.
The $\cPg\cPg \to \cPZ\cPZ $ contribution is estimated using events
generated with the {\tt gg2ZZ}
 code~\cite{Binoth:2008pr}.
The Z+jets samples, namely $\cPZ\cPqb\cPaqb$ , $\cPZ\cPqc\cPaqc$, and $\cPZ$+light jets,
are generated with \MADGRAPH. The $\ttbar$ events are generated at NLO with
\POWHEG.
For leading-order generators, the default set of parton distribution functions
(PDFs) used to produce these samples is \textsc{cteq6l}~\cite{CTEQ66}, while
\textsc{CT10}~\cite{ct10} is used for NLO generators.
Finally, for the
modeling of ATGCs, the \SHERPA generator version 1.2.2 is used~\cite{Sherpa}.
The $\Pgt$-lepton decays are generated with \TAUOLA~\cite{Jadach:1993hs}.
All events are processed through a detailed simulation of the CMS detector based on
\GEANTfour~\cite{GEANT} and reconstructed with the same algorithms as used for data.

A detailed description of the CMS detector can be found elsewhere~\cite{CMSExperiment}.
The central feature of the CMS apparatus
is a superconducting solenoid of 6\unit{m} internal diameter, providing
a magnetic field of 3.8\unit{T}. Within the field volume are the silicon pixel
and strip tracker, the crystal electromagnetic calorimeter (ECAL),
and the brass/scintil\-lator hadron calorimeter (HCAL).
Muons are measured in gas-ionization detectors embedded
in the steel flux return yoke of the magnet.
CMS uses a right-handed coordinate system, with the
origin at the nominal interaction point, the $x$ axis
pointing to the center of the LHC ring, the $y$ axis pointing
up (perpendicular to the plane of the LHC ring), and the $z$ axis
along the counterclockwise-beam direction. The polar angle
$\theta$ is measured from the positive $z$ axis and the
azimuthal angle $\phi$ is measured in the $x$-$y$ plane.
Variables used in this analysis include
the pseudorapidity
$\eta = -\ln[\tan(\theta/2)]$ and the transverse
momentum $\PT = \sqrt{p_x^2 + p_y^2}$.
The ECAL is designed to have both excellent energy resolution
and high granularity, properties that are crucial for reconstructing
electrons and photons produced in $\Pgt$-lepton decays.
The ECAL is constructed with projective lead tungstate
crystals that
provide coverage in pseudorapidity $\vert \eta \vert< 1.48 $ in a barrel region and
$1.48 <\vert \eta \vert < 3.00$ in two endcap regions (EE). A preshower detector consisting of two planes
of silicon sensors interleaved with a total of $3X_0$ of lead is located in front of the EE.
The energy resolution is 3\% or better for the range of electron energies relevant for this analysis.
The tracker measures charged particle tracks
within the range $|\eta| < 2.5$.  It consists of
$1440$ silicon pixel and $15\, 148$ silicon strip detector modules,
and provides an impact parameter resolution of
${\sim}15\micron$ and a transverse momentum
resolution of about 1.5\% for
100\GeV particles.
The reconstructed tracks are used to measure the location of interaction vertices.
The spatial resolution of the reconstruction in the transverse direction is
${\sim}25\micron$ for primary vertices with more than 30 associated tracks~\cite{tracker}.
The barrel region of the muon system is instrumented
with drift tubes, and the endcap regions with cathode
strip chambers.  In both regions, resistive-plate
chambers provide additional coordinate and timing
information.  Muons are reconstructed in the
range $|\eta| < 2.4$, with a typical $\PT$
resolution of ${\sim}1\%$ for $\PT = 40\GeV$.

At the trigger level, the selected events are
required to have either at least two electrons, one
with $\pt > 17\GeV$ and the other with $\pt > 8\GeV$,
or at least two muons, one
with $\pt > 13\GeV$ ($\pt > 17\GeV$ for
high instantaneous luminosity data-taking periods)
 and the other with $\pt > 8\GeV$.

Electrons are reconstructed within $|\eta^{\Pe}| < 2.5$ and with $\PT^{\Pe} > 7\GeV$
by combining information from the ECAL and tracker~\cite{Baffioni:2006cd,CMS-PAS-EGM-10-004}.
Electron identification requirements rely on the electromagnetic shower shape
and other observables based on tracker and calorimeter information.
The selection criteria depend on $\PT^{\Pe}$ and $| \eta^{\Pe} |$, and on a categorization
according to observables that are sensitive to the amount of bremsstrahlung emitted along the trajectory
in the tracker.
Muons are reconstructed~\cite{CMS-PAS-MUO-10-002} within $|\eta^{\Pgm}| < 2.4$ and $\PT^{\Pgm} > 5\GeV$
with information from both the tracker and the muon spectrometer.
The track must have more than 10 out of up to 24 possible
 hits in the silicon tracker~\cite{tracker}
to ensure a precise  measurement of the momentum.
The efficiencies are measured in data, using a tag-and-probe technique~\cite{CMS:2011aa}
based on an inclusive sample of $\cPZ \to \ell^+\ell^-$  events.
The measurements are performed in several ranges of $\PT^{\ell} $ and $ |\eta^{\ell}| $.
The product of reconstruction and identification efficiencies for electrons in the ECAL barrel (endcaps)
varies from about  68\% (62\%) for the  $\PT^{\Pe}$ range 7--10\GeV,
to  82\% (74\%) at $10 < \PT^{\Pe} < 20\GeV$,
and reaches up to 90\% (89\%) at $\PT^{\Pe} > 20\GeV$.
The muons are reconstructed and identified with efficiencies above $98\%$.

Since the $\PZ\PZ$ final state is expected to
have only a small contribution from background processes, the algorithms are
tuned to maximize the lepton-reconstruction efficiency,
resulting in an increased lepton-misidentification rate.
A particle-flow (PF) technique~\cite{CMS-PAS-PFT-09-001} is used
for $\tauh$ reconstruction.
In the
PF approach, information from all subdetectors is combined
to reconstruct and identify particles produced in the collision.
The particles are classified into mutually exclusive
categories:
charged hadrons, photons, neutral hadrons, muons, and
electrons.
These particles are used to reconstruct
the $\tauh$ candidates
with
the ``hadron plus strip'' (HPS) algorithm~\cite{Chatrchyan:2011xq}, which
is designed to optimize the performance of
$\tauh$ identification and reconstruction
by considering specific $\tauh$
decay modes.
The neutrinos produced in all $\Pgt$ decays escape detection
and are ignored in the $\tauh$ reconstruction.
The algorithm provides high $\tauh$ identification efficiency,
approximately 50\% for the range of $\tauh$ energies relevant for this analysis,
while keeping the misidentification rate for jets at the level
of 1\%.

Events are required to have at least one
$\cPZ \to \ell^+\ell^-$ candidate, denoted by $\cPZ_1$.
The invariant mass of the reconstructed $\cPZ_1$
is required to be
$ 60 < m_{\ell\ell} < 120\GeV$.
The two leptons must have opposite charges,
one
with
$\pt > 20\GeV$ and the other with $\pt > 10\GeV$,
and with
$|\eta| < 2.5$ for the electrons
and $|\eta| <  2.4$ for the muons.
If more than one candidate is found, the one with the mass closest
to the $\cPZ$ mass is considered as $\cPZ_1$.

Lepton isolation requirements depend on the $\cPZ\cPZ$ decay mode.
For the final states with only
electrons and muons,
the isolation criteria are based on a combination of
the tracker, ECAL, and
HCAL information.
The standard combined relative isolation is defined as

\begin{equation*}
I_\text{rel}^{\mathrm{std}} = \left( \sum_{i}
p_{\text{T, track}}^i + \max(\sum_{j}  E_{\text{T, ECAL}}^j
+ \sum_{k}  E_{\text{T, HCAL}}^k - \pi \cdot \Delta R^2_{\max} \cdot \rho ; 0) \right)/\pt^{\ell},
\end{equation*}

with the sums running over the charged tracks and the energy deposits
in the ECAL and HCAL within a cone around the lepton direction
defined by  $\Delta R  < \Delta R_{\rm max} = 0.3$,
where
$\Delta R = \sqrt { (\Delta \eta)^2 + (\Delta \phi)^2 }$, and $\ET$ stands for the
transverse energy.
The neutral isolation is made largely independent of the pileup of pp collisions
by correcting for the average energy
density, $\rho$,
calculated in each event using a ``jet area'' technique~\cite{rhocorrection}
and defined as the median of the energy distribution for the neutral particles around all
jets.
The isolation variable $I_\text{rel}^{\text{std}}$ is
required to be less than 0.275
for each lepton.
The significance of the impact parameter of each lepton relative
to the event vertex
($S_{\text{3D}}$)
is required to  satisfy  $| S_{\text{3D}} | < 4$.
The
primary vertex is chosen as the vertex with the highest sum of $\PT^2$
of its constituent tracks.

In the $\taus$ final states, instead of standard isolation,
the leptons from the $\cPZ_1$ are required to have
a combined PF relative isolation $I_\text{rel}^\mathrm{PF} < 0.25$.
The $I_\text{rel}^\mathrm{PF}$ is defined similarly to
$I_\text{rel}^{\mathrm{std}}$, however in this case the sums run over
charged hadrons, photons, and neutral hadrons,
all measured in the isolation cone of
$\Delta R < 0.4$
around the lepton direction.

The selection requirements for the second $\cPZ$,
denoted by $\cPZ_2$, also depend on the final state.
In 
the final states with electrons and muons only, the isolation requirements are the same as for the leptons
from $\cPZ_1$, but
$\pt > 7\GeV$ and $\pt > 5\GeV$ are required for electrons and muons, respectively.
If the final state is $\taue \taumu$,
the lepton $\pt$ values are required
to exceed 10\GeV. The remaining criteria are
identical to those
for $\cPZ_1$.
Since hadronically decaying $\Pgt$ leptons have much larger misidentification rates
than the other leptons,
the isolation requirement based on $I_\text{rel}^\mathrm{PF}$ for the electrons and
muons in the final states $\taue \tauh$ and $\taumu \tauh$ is changed to
0.15 and 0.1, respectively.
A study of inclusive $\cPZ \to \Pgt^+\Pgt^-$ production~\cite{CMS-EWK-TAU} demonstrated
that modifying the electron and muon
isolation requirements is  a more effective way to reduce background in such final
states than
requiring tighter isolation on $\tauh$.
The $\Pgt$ leptons are required to have $\pt > 20\GeV$ and $|\eta| <  2.3$, and to
satisfy the requirements of a loose
HPS working point. If the $\cPZ_2$ decays to $\tauh^+ \tauh^-$,
both $\tauh$ are required to satisfy the requirements of a medium working point of the HPS algorithm.
The loose (medium) working point requires the scalar sum of the
{\pt} of the charged
hadrons and {\ET} of the neutral hadrons within the isolation cone
to be less
than 2\GeV (1\GeV).
The loose (medium) working point corresponds to a probability of approximately
1\% (0.5\%) for jets to be misidentified as $\tauh$. Using the medium
instead of loose
working
point
leads to a decrease in the
$\tauh$ reconstruction efficiency
from
${\approx}50\%$ to ${\approx}40\%$.

The invariant mass of the reconstructed $\cPZ_2$
is required to satisfy
$ 60 < m_{\ell\ell} < 120\GeV$, when $\cPZ_2$ decays into $\Pep\Pem$ or $\Pgmp\Pgmm$.
In the $\taus$ final states,
the visible invariant mass of the
reconstructed $\cPZ_2 \to \Pgt^+\Pgt^-$ is required to satisfy
$ 30 < m^{\rm vis}_{\Pgt\Pgt} < 80\GeV$.
The upper bound
reduces contributions
from $\cPZ_2 \to \ell^+\ell^-$, where an electron or a muon is not well reconstructed and is
misidentified as a $\tauh$.
For the $\cPZ_2 \to \taue\taumu$ final state, the upper bound on
$m_{\Pgt\Pgt}$ is increased to 90\GeV, as this
state is not produced in $\cPZ_2 \to \ell^+\ell^-$ decays.
In the final states involving $\tauh$,
leptons from the same $\cPZ$ are required to be
separated by $\Delta R > 0.4$ for the $\cPZ_1$, and
by $\Delta R > 0.5$ for the $\cPZ_2$.

The major contributions to the background are due to
$\cPZ$ production in association with jets,
$\PW\cPZ$ production in association with
jets, and
$\ttbar$.
In all of these cases, a
jet or nonisolated lepton is misidentified as an isolated
electron, muon, or $\tauh$.
The relative contribution of each source of background
depends on the final state.

The background estimate is
performed in two steps. Firstly, the rate for
loosely isolated objects to be misidentified as isolated ones
is measured in a control region that does not contain any signal
contribution.
The misidentification rate
is estimated with events in which the
$\cPZ_1$ passes all selection requirements, and which contain
an additional probe electron, muon, or $\tauh$. No isolation
requirement is applied to the probe.
The misidentification rate is defined as the ratio of the number of
probe candidates that pass
the isolation requirements
to the initial number of probe candidates, and  is measured as a function of
$\pt$ and $\eta$ for each lepton flavor.

The second step is to estimate the number of background events in the signal region.
The
measured misidentification rate
is applied
to  events that  pass all selection requirements, including
the opposite-charge requirement for the $\PZ_2$,
but requiring
the candidates to not be isolated.

Theoretical uncertainties on the $\cPZ\cPZ \to 2\ell 2\ell'$ acceptance are evaluated using
\textsc{mcfm}\xspace6.2~\cite{MCFM},
varying QCD scales up and down by a factor of two with respect to the
default factorization ($\mu_{\mathrm{F}}$) and renormalization ($\mu_{\mathrm{R}}$) scales
$\mu_{\mathrm{F}} = \mu_{\mathrm{R}} = m_\cPZ$, where $m_\cPZ$ is the mass of the $\cPZ$
boson. The variations in the acceptance are 0.1\% ($\cPq\cPaq \to \cPZ\cPZ$) and
0.4\% ($\cPg\cPg \to \cPZ\cPZ$) and can be neglected.
The uncertainties related to the PDFs
are evaluated following the PDF4LHC prescription~\cite{Botje:2011sn}.
Using the CT10~\cite{ct10}, MSTW08~\cite{Martin:2009iq}, and NNPDF~\cite{nnpdf} sets, the uncertainties are estimated to be 4\%
for $\cPq\cPaq \to \cPZ\cPZ$ and
and 5\% for
$\cPg\cPg \to \cPZ\cPZ$ processes.

The uncertainties on \cPZ+jets, \PW\cPZ+jets, and $\ttbar$ backgrounds reflect
the uncertainties on the measured values of the misidentication rates and the
limited quantity of data in the control regions in the data and amount to  30--50\%
depending on the decay channel.
The uncertainty on the integrated luminosity is 2.2\%~\cite{lumiPAS12}.
Systematic uncertainties on trigger efficiency (1.5\%), lepton
identification efficiency, and lepton isolation are evaluated from data.
The uncertainties associated with lepton
identification and isolation are 1--2\% for muons and electrons, and 6--7\%
for $\tauh$.
Uncertainties on energy scales, 3\% for $\tauh$ and 1--2.5\% for electrons, contribute
to variations in the shape of the mass spectrum.

\begin{table}[htbp]
\centering
\topcaption{ The expected yield of $\cPZ\cPZ$ events obtained from simulation and
the estimated yield of background
events obtained from data,
as described in the text, are shown for each decay channel and
are summed in the total expected yield (``Total expected'').
They are compared to the number of events observed in the signal region.
The first uncertainty is
statistical while the second one is systematic.
}
\label{table:results}
\begin{tabular}{|c|c|c|c|c|}
\hline
Decay & Expected $\PZ\PZ$ &  Background   & Total & Observed \\
channel &  &  & expected  & \\
\hline
$\Pe\Pe\Pe\Pe$ 			& $ 10.50 \pm 0.04 \pm 0.95 $ & $ 0.25 \pm 0.14 \pm 0.07 $ & $10.75 \pm 0.14 \pm 0.95$ & $  9 $\\
$\Pgm\Pgm\Pgm\Pgm$ 		& $ 15.91 \pm 0.05 \pm 1.43 $ & $ 0.52 \pm 0.26 \pm 0.25 $ & $16.43 \pm 0.26 \pm 1.45$ & $ 14 $\\
$\Pe\Pe\Pgm\Pgm$ 		& $ 26.74 \pm 0.10 \pm 2.41 $ & $ 0.58 \pm 0.18 \pm 0.23 $ & $27.32 \pm 0.21 \pm 2.41$ & $ 31 $\\
\hline
$\Pe\Pe\tauh\tauh$ 			& $ 0.75 \pm 0.01 \pm 0.07 $ & $ 0.76 \pm 0.16 \pm 0.05 $ & $1.51 \pm 0.16 \pm 0.09 $ & $ 1 $\\
$\Pgm\Pgm\tauh\tauh$ 			& $ 0.82 \pm 0.02 \pm 0.07 $ & $ 0.75 \pm 0.16 \pm 0.08 $ & $1.57 \pm 0.16 \pm 0.11 $ & $ 0 $\\
$\Pe\Pe\Pgt_{\Pe}\tauh$ 		& $ 1.17 \pm 0.02 \pm 0.11 $ & $ 0.96 \pm 0.34 \pm 0.12 $ & $2.29 \pm 0.34 \pm 0.16 $ & $ 3 $\\
$\Pgm\Pgm\Pgt_{\Pe}\tauh$ 		& $ 1.15 \pm 0.02 \pm 0.10 $ & $ 0.35 \pm 0.34 \pm 0.11 $ & $1.60 \pm 0.34 \pm 0.15 $ & $ 3 $\\
$\Pe\Pe\Pgt_{\Pgm}\tauh$ 		& $ 0.94 \pm 0.02 \pm 0.08 $ & $ 0.22 \pm 0.14 \pm 0.04 $ & $1.17 \pm 0.14 \pm 0.06 $ & $ 0 $\\
$\Pgm\Pgm\Pgt_{\Pgm}\tauh$ 		& $ 1.08 \pm 0.02 \pm 0.10 $ & $ 0.55 \pm 0.24 \pm 0.11 $ & $1.64 \pm 0.24 \pm 0.15 $ & $ 2 $\\
$\Pe\Pe\Pgt_{\Pe}\Pgt_{\Pgm}$ 		& $ 0.54 \pm 0.01 \pm 0.05 $ & $ 0.64 \pm 0.44 \pm 0.16 $ & $1.22 \pm 0.44 \pm 0.17 $ & $ 0 $\\
$\Pgm\Pgm\Pgt_{\Pe}\Pgt_{\Pgm}$ 	& $ 0.60 \pm 0.01 \pm 0.05 $ & $ 0.14 \pm 0.30 \pm 0.10 $ & $0.74 \pm 0.30 \pm 0.11 $ & $ 2 $\\
\hline
\end{tabular}
\end{table}

\begin{figure}[htbp]
\begin{center}
\includegraphics[width=0.45\textwidth]{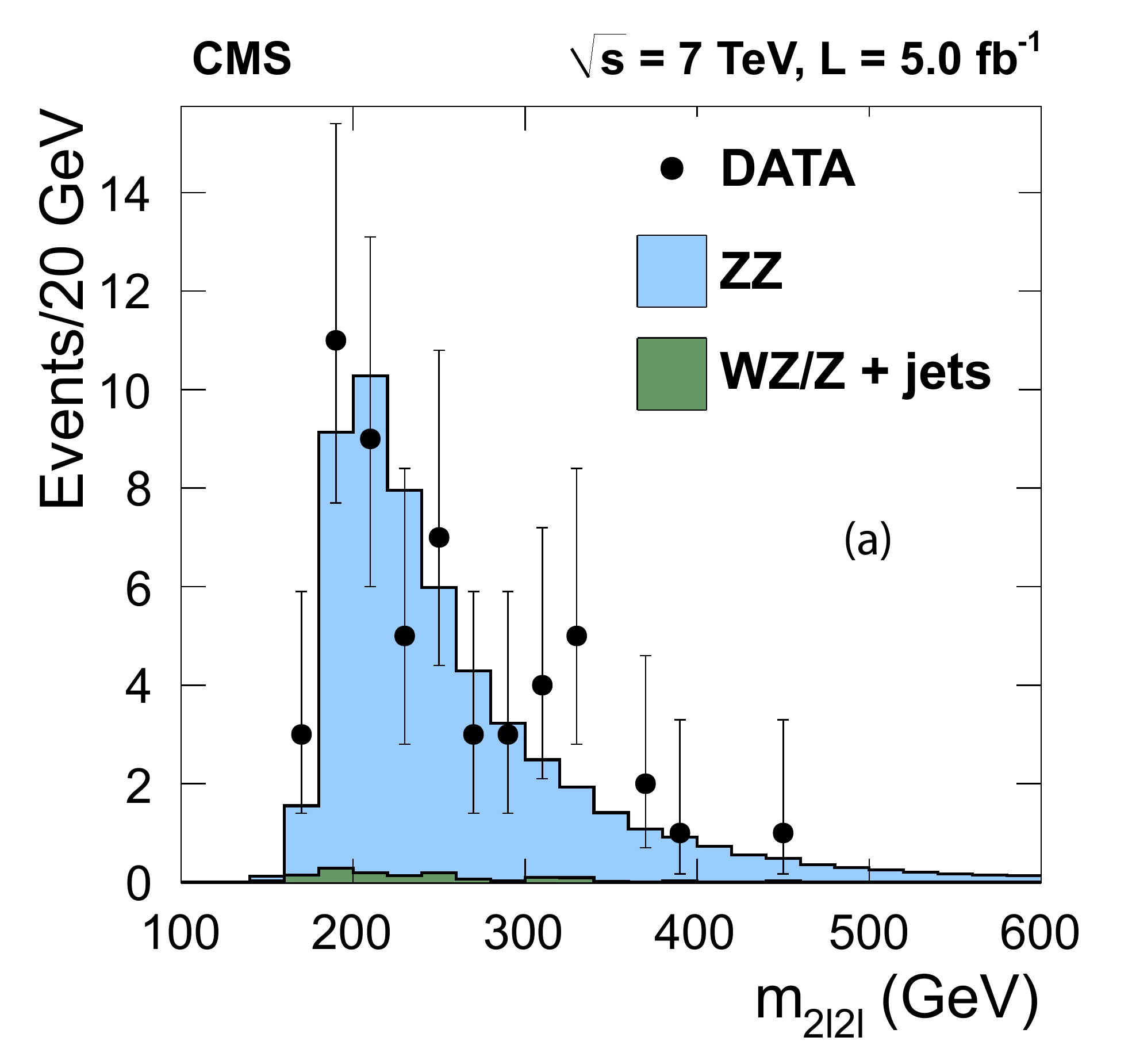}
\includegraphics[width=0.45\textwidth]{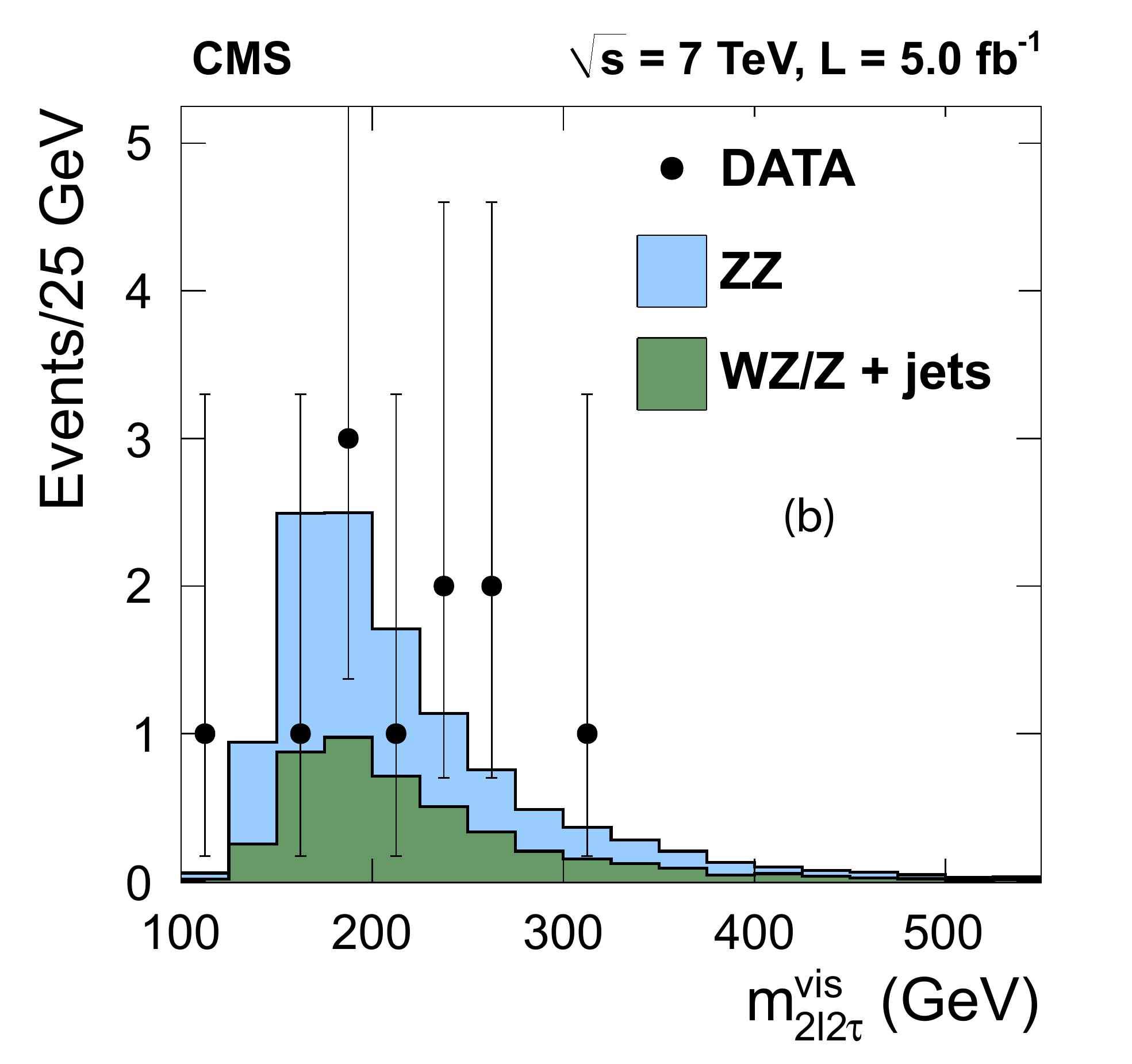}
\includegraphics[width=0.45\textwidth]{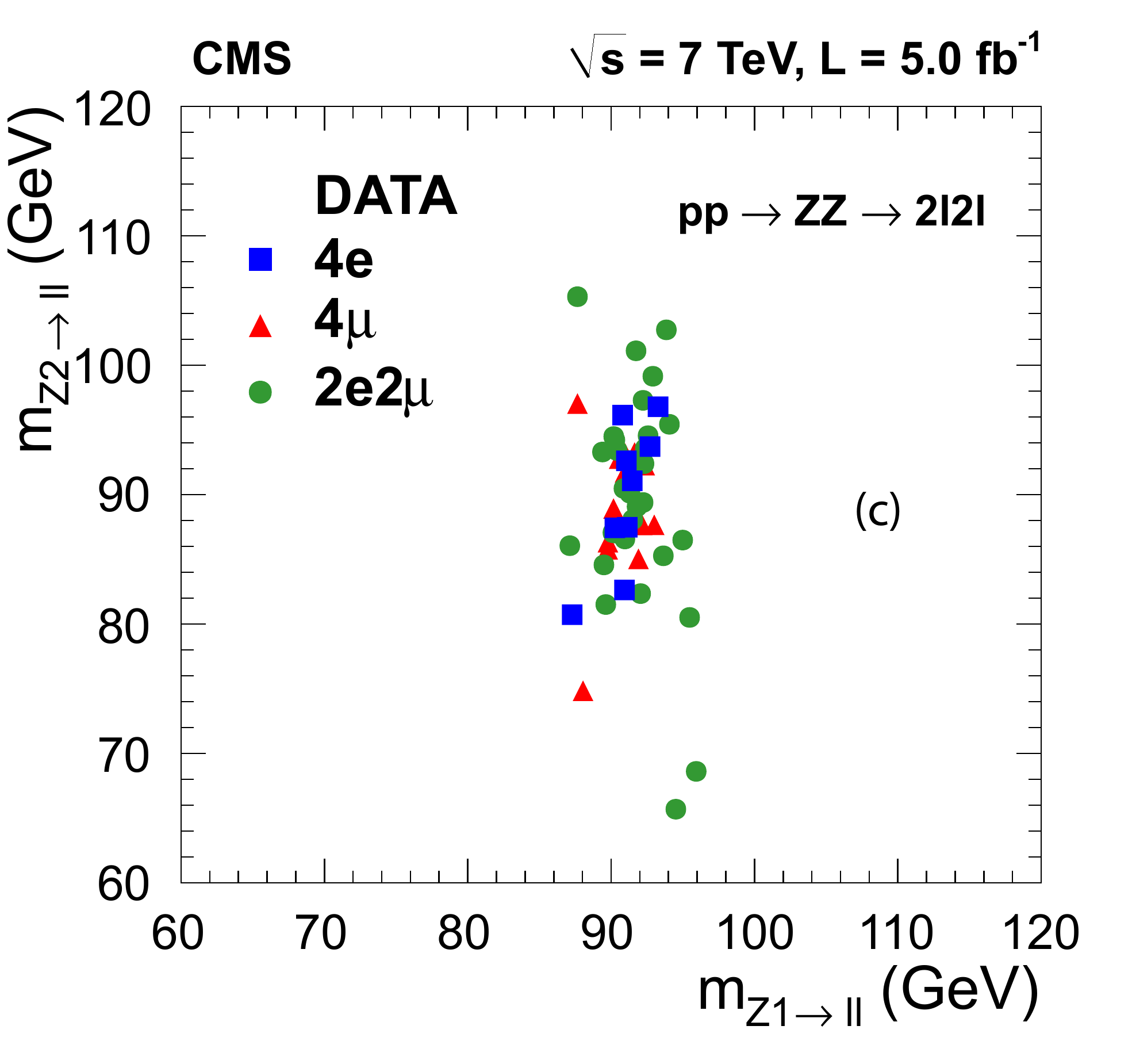}
\includegraphics[width=0.45\textwidth]{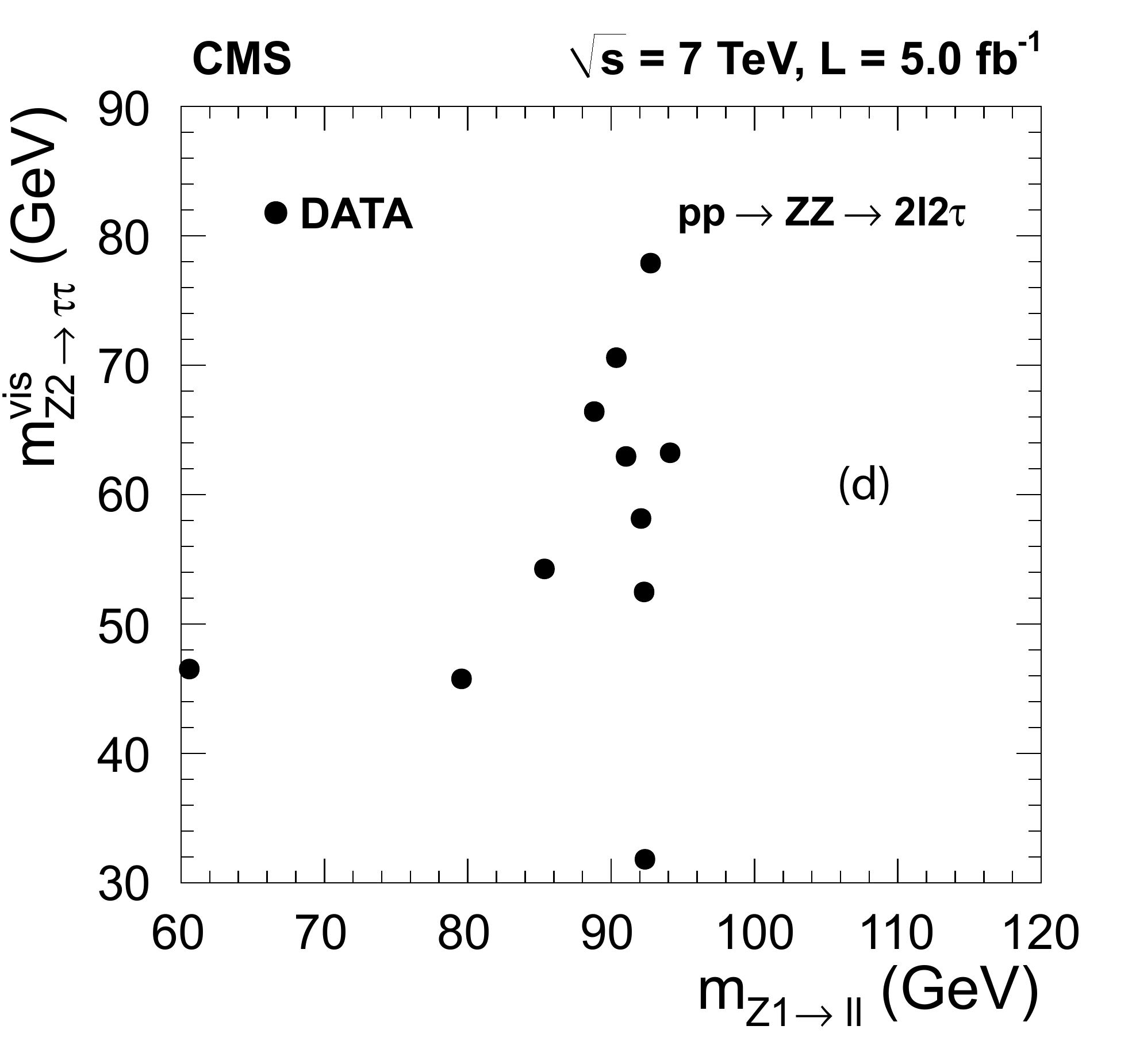}
\caption{Distributions of the four-lepton reconstructed mass for (a) the sum of the
$4\Pe, 4\Pgm$, and $2\Pe2\Pgm$ channels and (b) the sum of the $\taus$ channels.
Points represent the data, and the shaded histograms represent
the expected $\PZ\PZ$ signal and the reducible background.
The shapes of the signal and background are taken from the MC simulation, with each
component normalized to
the corresponding estimated value from Table~\ref{table:results}.
The distributions (c) and (d) demonstrate the relationship between
the reconstructed $\PZ_1$ and $\PZ_2$ masses.}
\label{fig:Mass4l}
\end{center}
\end{figure}

Table~\ref{table:results} presents the number of observed events
in the signal region in each channel, as well as
the expected number of signal events and the estimated
number of background events.
We observe a total of 54 candidate events
in the
$4\Pe, 4\Pgm$, and $2\Pe2\Pgm$
channels,
compared to the SM expectation of $54.5 \pm 0.3 \stat \pm 4.8 \syst$ events, which
includes 1.4 from background
processes.
In the $\taus$ channels,
11 candidate events are observed, compared
to  $11.7 \pm 0.8\stat \pm 1.0 \syst$ events expected, including
4.4 from background processes.
The reconstructed four-lepton invariant mass distributions
are compared
to the SM expectations
 in Figs.~\ref{fig:Mass4l} (a) and (b) for the sum of the
$4\Pe, 4\Pgm$, and $2\Pe2\Pgm$ channels, and the sum of
all the $\taus$ channels.
The shapes of the signal and background are taken from the MC simulation, with each
component normalized to
the corresponding estimated value from Table~\ref{table:results}.
The reconstructed masses in $\taus$ states ($m^{\rm vis}_{\taus}$) are shifted downwards
with respect to the generated $\cPZ$ masses by about $30\%$ due to the
undetected  neutrinos in $\tau$ decays.
Figures~\ref{fig:Mass4l} (c) and (d) demonstrate the relationship between
the reconstructed $\PZ_1$ and $\PZ_2$ masses.

To include all the final states in the calculation of the
cross section,
a simultaneous fit to the numbers of  observed events in all the
decay channels
is performed. The fit is constrained by the requirement that all the  measurements
come from the same initial state via different decay modes.
It allows for combining  many decay modes with either
very few or no events observed.
The joint likelihood is a combination of the likelihoods for the
individual channels, which include
the signal and background hypotheses.
Each $\Pgt$-lepton decay mode is treated
as a separate channel
because they are mutually exclusive owing to the
methodology adopted for event
reconstruction and subsequent event selection.
The statistical and systematic uncertainties are introduced in the form of
nuisance parameters
via log-normal distributions around the estimated central values.

The resulting cross section is measured to be

\begin{equation*}
\sigma ( \Pp\Pp \to \PZ\PZ) =  6.24 \,^{+0.86}_{-0.80} \stat \,^{+0.41}_{-0.32} \syst \pm 0.14 \lumi \unit{pb}.
\end{equation*}

This result is
to be compared to the theoretical value of
${\rm 6.3 \pm 0.4\, pb}$
 calculated with {\textsc{MCFM}\xspace}
at NLO for $\cPq\cPaq \to \cPZ\cPZ$ and
LO for $\cPg\cPg \to \cPZ\cPZ$
with the
MSTW08 PDFs and for both $\cPZ$ bosons in the mass range $60 < M_{\PZ} < 120\GeV$.
This is the most precise published
$\Pp\Pp \to \cPZ\cPZ$ cross section
measurement to date, which for the first time extends the $\Pp\Pp \to \cPZ\cPZ \to 2\ell2\ell$ measurement to include
final states with hadronically decaying $\Pgt$ leptons.

\begin{figure}[htbp]
\begin{center}
\includegraphics[width=0.45\textwidth]{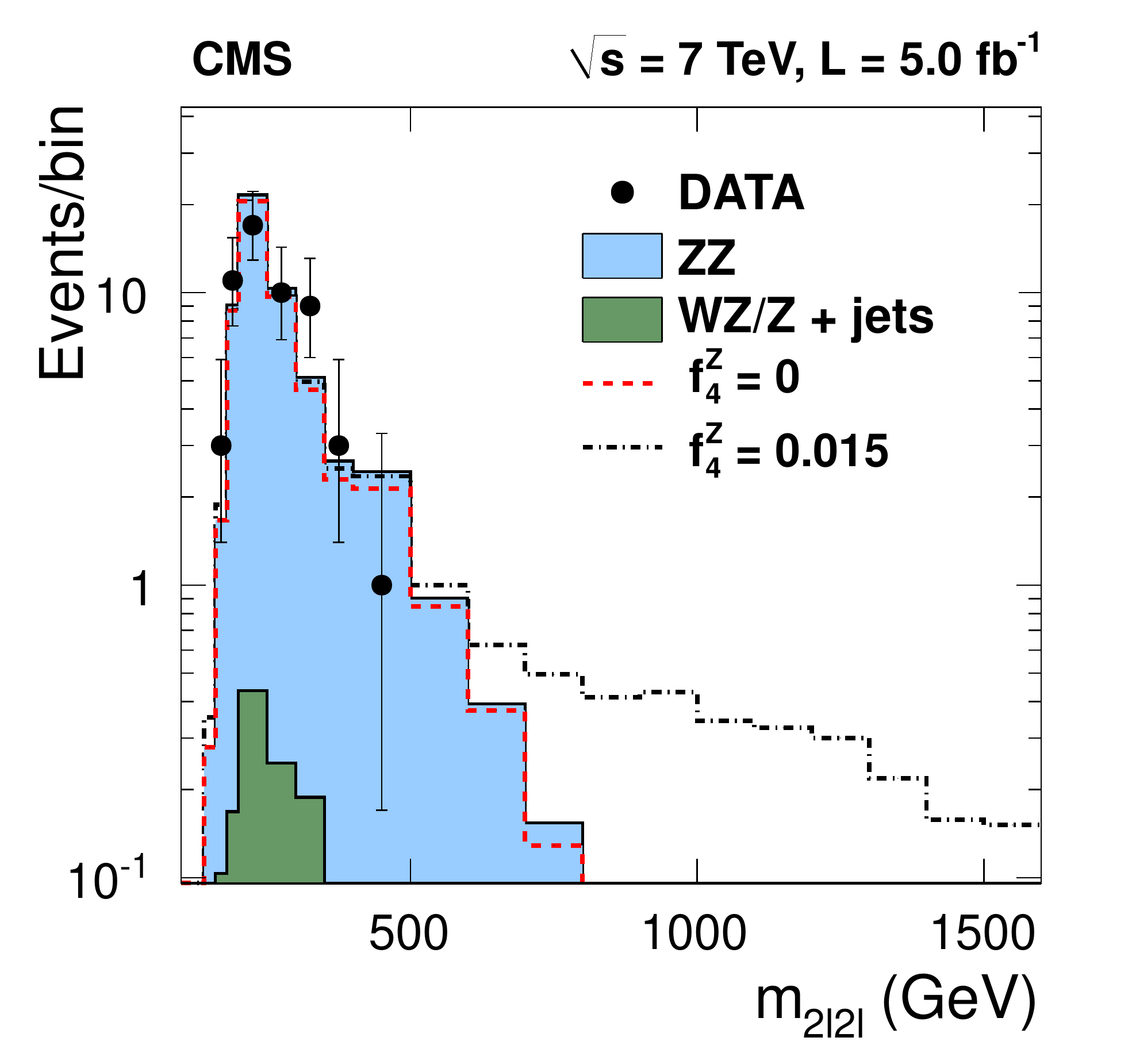}
\caption{Distribution of the four-lepton reconstructed mass for the sum of the
$4\Pe, 4\Pgm$, and the $2\Pe2\Pgm$ channels.
Points represent the data, and the shaded histograms represent
the expected $\PZ\PZ$ signal and the reducible background.
The dashed and dotted histograms represent the results of the \SHERPA
simulation for the SM ($f_4^{\cPZ} = 0$) and
 in the presence of an ATGC ($f_4^{\cPZ}=0.015$), while
all the other anomalous couplings are set to zero.
 }
\label{figure:sherpa4l}
\end{center}
\end{figure}

\begin{figure}[htbp]
\begin{center}
\includegraphics[width=0.45\textwidth]{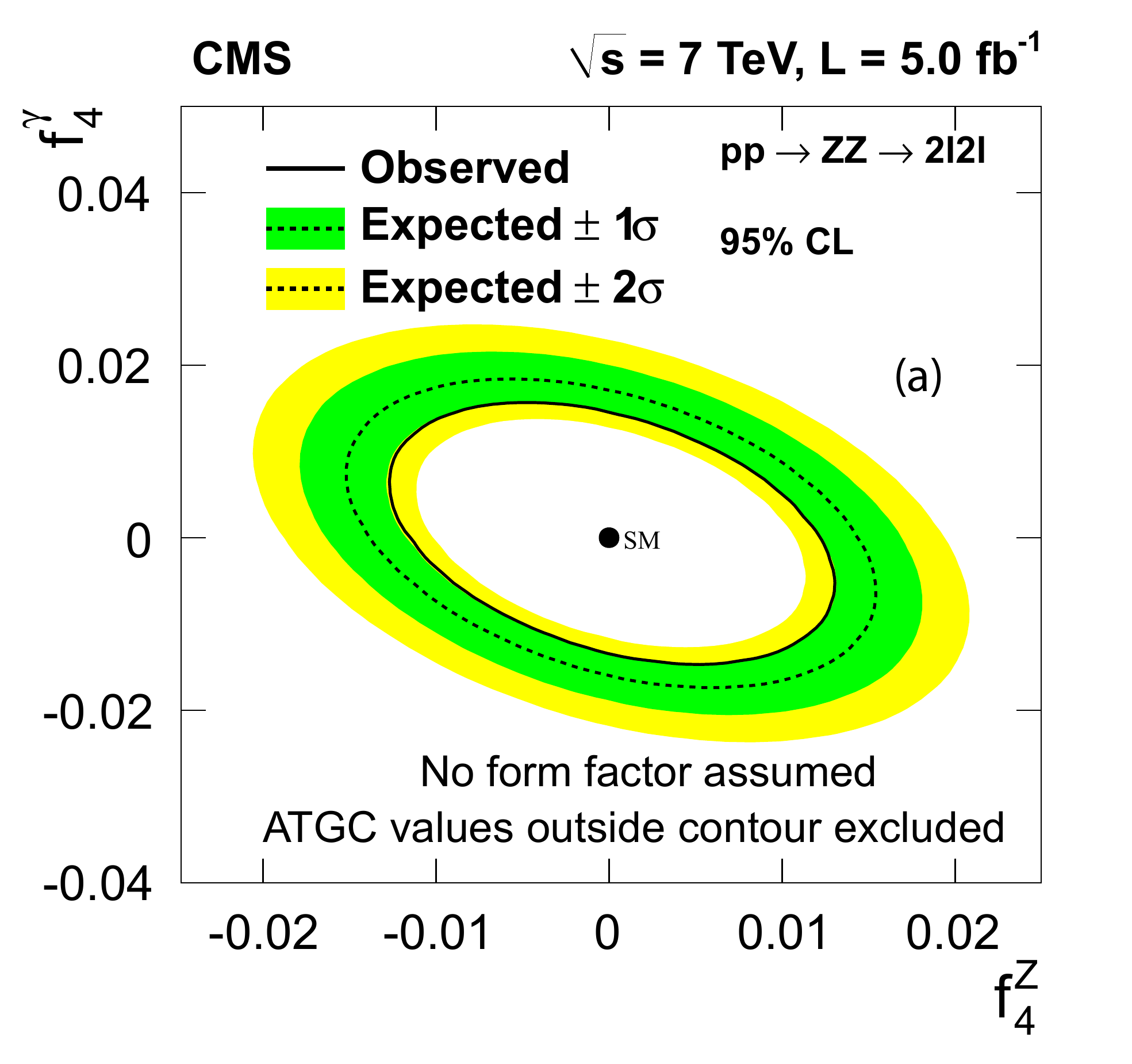}
\includegraphics[width=0.45\textwidth]{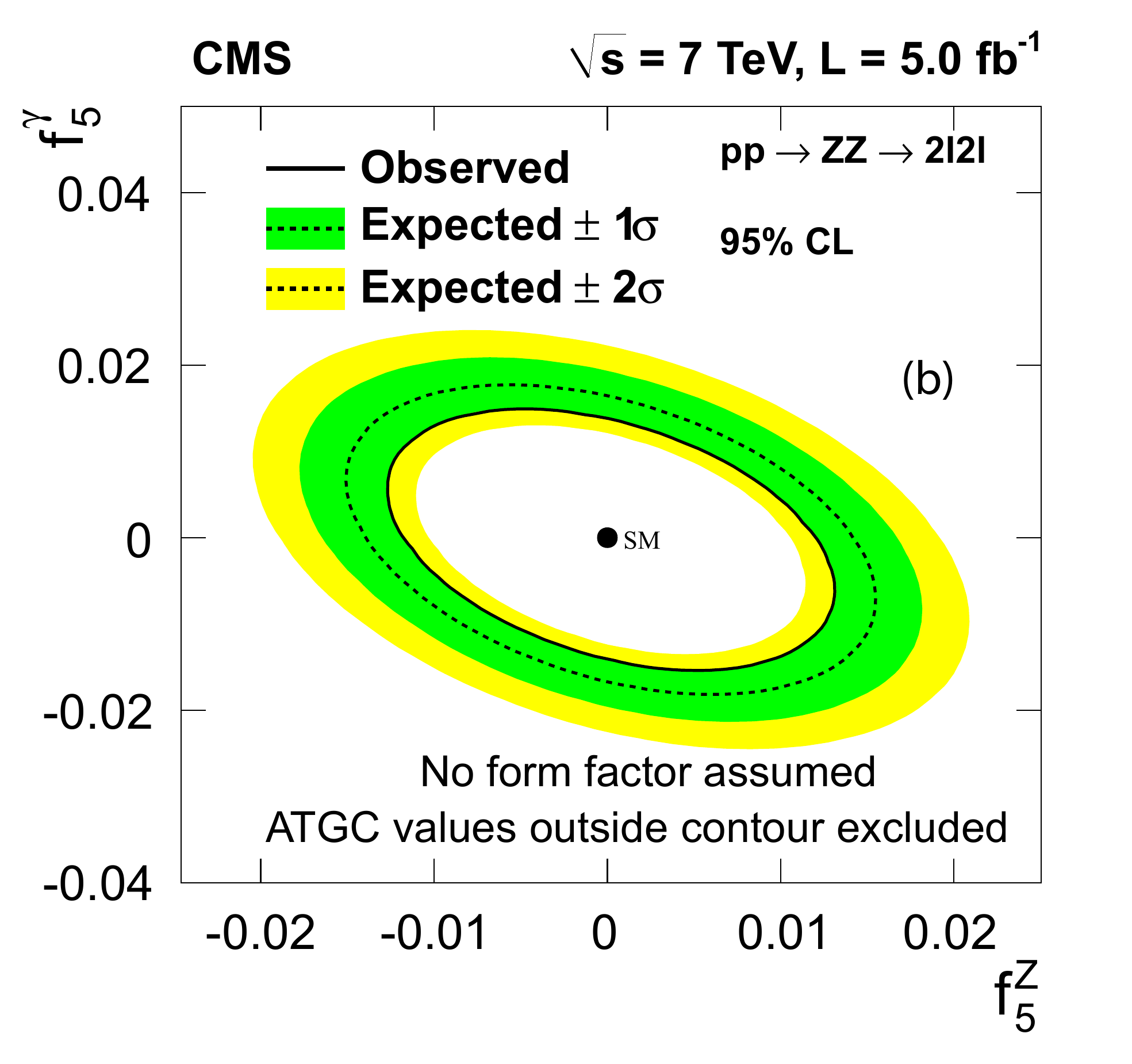}
\caption{Expected
and observed two-dimensional exclusion limits at 95\% CL on the
anomalous neutral trilinear $\cPZ\cPZ\cPZ$ ($f_{4,5}^{\cPZ}$) and
$\cPZ\cPZ\gamma$ ($f_{4,5}^{\gamma}$) couplings. The green and yellow bands
represent the one and two standard-deviation variations from the expected limit.
In calculating the limits, the anomalous couplings that are not shown in the figure are
set to zero.}
\label{figure:limitPlot}
\end{center}
\end{figure}

The limits on ATGCs are calculated with the modified frequentist
construction CL$_{\mathrm{s}}$~\cite{CLS1, CLS2, ATL-PHYS-PUB-2011-011} based on the shape of the
four-lepton invariant mass distributions, including the
$4\Pe$, $4\Pgm$, and 	
$2\Pe2\Pgm$ channels
in the likelihood
combination.
Figure~\ref{figure:sherpa4l} presents the
distribution of the four-lepton reconstructed mass for the sum of the
$4\Pe, 4\Pgm$, and $2\Pe2\Pgm$ channels.
The dashed and dotted histograms represent the results of the \SHERPA
simulation for the SM ($f_4^{\cPZ} = 0$) and
 in the presence of an ATGC ($f_4^{\cPZ}=0.015$), while
all the other anomalous couplings are set to zero.
The presence of ATGCs
would be manifested in an increased yield of events at high four-lepton masses.
The invariant mass distributions are interpolated from
\SHERPA simulation for different values of the anomalous couplings. For
each distribution only one or two couplings are varied, while all others are set to zero.
The fit is performed
to find the maximum likelihood value and limits are calculated.
To avoid unitarity violation at energies above the scale
 $\Lambda$ of new physics,
the ATGCs are often modified with a form-factor parametrization
of the type
 $1/(1+\hat{s}/\Lambda^2)^2$, where
$\sqrt{\hat{s}}\approx m_{2\ell2\ell}$ is
the effective center-of-mass energy of the collision.
However, no unitarity violations occur in the sensitive
region $m_{2\ell2\ell} \lesssim 1.5\TeV$ for
bare anomalous couplings of order $0.05$ or smaller~\cite{Zralek:1989ty},
so we calculate the limits
 without form-factor scaling.
This choice has the advantage of avoiding
any bias from energy-dependence assumptions and is exact
in the limit in
which the scale of new physics is much larger than
$\sqrt{\hat{s}}$.

Figure~\ref{figure:limitPlot} presents the
expected and observed two-dimensional exclusion limits at 95\% confidence level (CL) on the
anomalous neutral trilinear $\cPZ\cPZ\cPZ$ and $\cPZ\cPZ\gamma$ couplings.
The green and yellow bands represent the one and two standard-deviation variations from the expected limit.
The present limits are dominated by statistical uncertainties. Systematic uncertainties arising from the uncertainty on the theoretical
cross section, PDFs, detector efficiencies, and luminosity are introduced in the form of nuisance
parameters with log-normal probability density functions.
One-dimensional 95\% CL limits
for the $f_4^{\cPZ,\gamma}$ and $f_5^{\cPZ,\gamma}$ anomalous coupling parameters are measured
to be

\begin{equation}
-0.011 < f_4^\cPZ< 0.012,\,
-0.012 < f_5^\cPZ < 0.012,\, -0.013 < f_4^\gamma < 0.015,\, -0.014 < f_5^\gamma < 0.014.
\nonumber
\end{equation}

In the one-dimensional fits, all of the ATGC
parameters except the one under study are kept fixed to zero.
These values extend previous results on vector boson self-interactions
and are currently the most stringent limits established for $\PZ\PZ\PZ$ and $\PZ\PZ\gamma$ couplings.

In summary, we have presented an updated measurement of the $\cPZ\cPZ$ production
cross section
in proton-proton collisions at 7\TeV in the $\cPZ \cPZ \to 2\ell 2\ell'$ decay mode,
with $\ell = \Pe, \Pgm$ and $\ell' = \Pe, \Pgm, \Pgt$.
The data sample corresponds to an integrated luminosity of $5.0\fbinv$.
The measured cross section
$\sigma ( \Pp\Pp \to \cPZ\cPZ) =  6.24 \,^{+0.86}_{-0.80} \stat \,^{+0.41}_{-0.32} \syst
\pm 0.14 \lumi \unit{pb}$
is consistent with the SM prediction and is the most precise published
$\Pp\Pp \to \cPZ\cPZ$ cross section measurement to date.
For the first time the $\Pp\Pp \to \PZ\PZ \to 2\ell2\ell$ measurements are extended to
include
final states with hadronically decaying $\Pgt$ leptons.
Limits on vector-boson self-interactions are established,
significantly restricting anomalous $\PZ \PZ \PZ$ and $\PZ \PZ \gamma$ trilinear gauge
couplings.

We congratulate our colleagues in the CERN accelerator departments for the excellent performance of the LHC and thank the technical and administrative staffs at CERN and at other CMS institutes for their contributions to the success of the CMS effort. In addition, we gratefully acknowledge the computing centers and personnel of the Worldwide LHC Computing Grid for delivering so effectively the computing infrastructure essential to our analyses. Finally, we acknowledge the enduring support for the construction and operation of the LHC and the CMS detector provided by the following funding agencies: BMWF and FWF (Austria); FNRS and FWO (Belgium); CNPq, CAPES, FAPERJ, and FAPESP (Brazil); MEYS (Bulgaria); CERN; CAS, MoST, and NSFC (China); COLCIENCIAS (Colombia); MSES (Croatia); RPF (Cyprus); MoER, SF0690030s09 and ERDF (Estonia); Academy of Finland, MEC, and HIP (Finland); CEA and CNRS/IN2P3 (France); BMBF, DFG, and HGF (Germany); GSRT (Greece); OTKA and NKTH (Hungary); DAE and DST (India); IPM (Iran); SFI (Ireland); INFN (Italy); NRF and WCU (Republic of Korea); LAS (Lithuania); CINVESTAV, CONACYT, SEP, and UASLP-FAI (Mexico); MSI (New Zealand); PAEC (Pakistan); MSHE and NSC (Poland); FCT (Portugal); JINR (Armenia, Belarus, Georgia, Ukraine, Uzbekistan); MON, RosAtom, RAS and RFBR (Russia); MSTD (Serbia); SEIDI and CPAN (Spain); Swiss Funding Agencies (Switzerland); NSC (Taipei); ThEPCenter, IPST and NSTDA (Thailand); TUBITAK and TAEK (Turkey); NASU (Ukraine); STFC (United Kingdom); DOE and NSF (USA).

\bibliography{auto_generated}   

\cleardoublepage \appendix\section{The CMS Collaboration \label{app:collab}}\begin{sloppypar}\hyphenpenalty=5000\widowpenalty=500\clubpenalty=5000\input{SMP-12-007-authorlist.tex}\end{sloppypar}
\end{document}

%% file: SMP-12-007-authorlist.tex
\textbf{Yerevan Physics Institute,  Yerevan,  Armenia}\\*[0pt]
S.~Chatrchyan, V.~Khachatryan, A.M.~Sirunyan, A.~Tumasyan
\vskip\cmsinstskip
\textbf{Institut f\"{u}r Hochenergiephysik der OeAW,  Wien,  Austria}\\*[0pt]
W.~Adam, E.~Aguilo, T.~Bergauer, M.~Dragicevic, J.~Er\"{o}, C.~Fabjan\cmsAuthorMark{1}, M.~Friedl, R.~Fr\"{u}hwirth\cmsAuthorMark{1}, V.M.~Ghete, J.~Hammer, N.~H\"{o}rmann, J.~Hrubec, M.~Jeitler\cmsAuthorMark{1}, W.~Kiesenhofer, V.~Kn\"{u}nz, M.~Krammer\cmsAuthorMark{1}, I.~Kr\"{a}tschmer, D.~Liko, I.~Mikulec, M.~Pernicka$^{\textrm{\dag}}$, B.~Rahbaran, C.~Rohringer, H.~Rohringer, R.~Sch\"{o}fbeck, J.~Strauss, A.~Taurok, W.~Waltenberger, G.~Walzel, E.~Widl, C.-E.~Wulz\cmsAuthorMark{1}
% ...
\vskip\cmsinstskip
\dag:~Deceased\\
1:~~Also at Vienna University of Technology, Vienna, Austria\\
% ...


%% file: SMP-12-007_temp.bbl
\providecommand{\href}[2]{#2}\begingroup\raggedright\begin{thebibliography}{10}%
\makeatletter
\providecommand{\hrefCMSnoop }[0]{\@secondoftwo}%
\makeatother
\providecommand{\doi}{\texttt{doi:}\begingroup \urlstyle{tt}\Url}

\bibitem{Hagiwara}
K.~Hagiwara\hrefCMSnoop {} { {et~al.}, ``Probing the weak boson sector in
  $\Pe^+\Pe^- \to \PW^+\PW^-$'',} \textit{ Nucl. Phys. B} \textbf{ 282} (1987)
  253,
\href{http://dx.doi.org/10.1016/0550-3213(87)90685-7}{\doi{10.1016/0550-3213(87)90685-7}}.

\bibitem{CDFZZxs}
\hrefCMSnoop {} {{ CDF} Collaboration, ``{Measurement of $\PZ\PZ$ production in
  leptonic final states at $\sqrt{s}$ of 1.96 TeV at CDF}'',} \textit{ Phys.
  Rev. Lett.} \textbf{ 108} (2012) 101801,
  \href{http://dx.doi.org/10.1103/PhysRevLett.108.101801}{\doi{10.1103/PhysRevLett.108.101801}},
\href{http://www.arXiv.org/abs/1112.2978}{\texttt{ arXiv:1112.2978}}.

\bibitem{D0ZZxs}
\hrefCMSnoop {} {{ D0} Collaboration, ``{Measurement of the $\cPZ\cPZ$
  production cross section in $\rm{p\bar{p}}$ collisions at
  $\sqrt{s}=1.96\TeV$}'',} \textit{ Phys. Rev. D} \textbf{ 84} (2011) 011103,
  \href{http://dx.doi.org/10.1103/PhysRevD.84.011103}{\doi{10.1103/PhysRevD.84.011103}},
\href{http://www.arXiv.org/abs/1104.3078}{\texttt{ arXiv:1104.3078}}.

\bibitem{HZZ4lPRL}
\hrefCMSnoop {} {{ CMS} Collaboration, ``Search for the standard model Higgs
  boson in the decay channel H$\rightarrow$ZZ$\rightarrow$4l in pp collisions
  at $\sqrt{s} = 7\TeV$'',} \textit{ Phys. Rev. Lett.} \textbf{ 108} (2012)
  111804,
  \href{http://dx.doi.org/10.1103/PhysRevLett.108.111804}{\doi{10.1103/PhysRevLett.108.111804}},
\href{http://www.arXiv.org/abs/1202.1997}{\texttt{ arXiv:1202.1997}}.

\bibitem{ATLAS}
\hrefCMSnoop {} {{ ATLAS} Collaboration, ``Measurement of the ZZ production
  cross section and limits on anomalous neutral triple gauge couplings in
  proton-proton collisions at $\sqrt{s} = 7\TeV$ with the ATLAS detector'',}
  \textit{ Phys. Rev. Lett.} \textbf{ 108} (2012) 041804,
  \href{http://dx.doi.org/10.1103/PhysRevLett.108.041804}{\doi{10.1103/PhysRevLett.108.041804}},
\href{http://www.arXiv.org/abs/1110.5016}{\texttt{ arXiv:1110.5016}}.

\bibitem{ALEPH}
\hrefCMSnoop {} {{ ALEPH} Collaboration, ``{Measurement of $\PZ$-pair
  production in $\Pe^+\Pe^-$ collisions and constraints on anomalous neutral
  gauge couplings}'',} \textit{ JHEP} \textbf{ 04} (2009) 124,
\href{http://dx.doi.org/10.1088/1126-6708/2009/04/124}{\doi{10.1088/1126-6708/2009/04/124}}.

\bibitem{DELPHI}
\hrefCMSnoop {} {{ DELPHI} Collaboration, ``{Study of triple-gauge-boson
  couplings ZZZ, ZZ$\gamma$ and Z$\gamma\gamma$ at LEP}'',} \textit{ Eur. Phys.
  J. C} \textbf{ 51} (2007) 525,
  \href{http://dx.doi.org/10.1140/epjc/s10052-007-0345-0}{\doi{10.1140/epjc/s10052-007-0345-0}},
\href{http://www.arXiv.org/abs/0706.2741}{\texttt{ arXiv:0706.2741}}.

\bibitem{LEPWG}
\hrefCMSnoop {} {{LEP Electroweak Working Group}, ``A combination of
  preliminary electroweak measurements and constraints on the standard
  model'',} (2006). \href{http://www.arXiv.org/abs/hep-ex/0612034}{\texttt{
  arXiv:hep-ex/0612034}}.

\bibitem{OPAL}
\hrefCMSnoop {} {{ OPAL} Collaboration, ``{Study of $\PZ$ pair production and
  anomalous couplings in $\Pe^+\Pe^-$ collisions at $\sqrt{s}$ between 190\GeV
  and 209\GeV}'',} \textit{ Eur. Phys. J. C} \textbf{ 32} (2003) 303,
  \href{http://dx.doi.org/10.1140/epjc/s2003-01467-x}{\doi{10.1140/epjc/s2003-01467-x}},
\href{http://www.arXiv.org/abs/hep-ex/0310013}{\texttt{ arXiv:hep-ex/0310013}}.

\bibitem{L3}
\hrefCMSnoop {} {{ L3} Collaboration, ``{Study of anomalous $\PZ\PZ \gamma$ and
  $\PZ \gamma \gamma$ couplings at LEP}'',} \textit{ Phys. Lett. B} \textbf{
  436} (1998) 187,
\href{http://dx.doi.org/10.1016/S0370-2693(98)00883-1}{\doi{10.1016/S0370-2693(98)00883-1}}.

\bibitem{D0}
\hrefCMSnoop {} {{ D0} Collaboration, ``{Search for $\PZ\PZ$ and $\PZ\gamma^*$
  production in $\rm{p\bar{p}}$ collisions at $\sqrt{s} = 1.96\TeV$ and limits
  on anomalous $\PZ\PZ\PZ$ and $\PZ\PZ\gamma^*$ couplings}'',} \textit{ Phys.
  Rev. Lett.} \textbf{ 100} (2008) 131801,
  \href{http://dx.doi.org/10.1103/PhysRevLett.100.131801}{\doi{10.1103/PhysRevLett.100.131801}},
\href{http://www.arXiv.org/abs/0712.0599}{\texttt{ arXiv:0712.0599}}.

\bibitem{Alioli:2008gx}
S.~Alioli\hrefCMSnoop {} { {et~al.}, ``{NLO vector-boson production matched
  with shower in POWHEG}'',} \textit{ JHEP} \textbf{ 07} (2008) 060,
  \href{http://dx.doi.org/10.1088/1126-6708/2008/07/060}{\doi{10.1088/1126-6708/2008/07/060}},
\href{http://www.arXiv.org/abs/0805.4802}{\texttt{ arXiv:0805.4802}}.

\bibitem{Nason:2004rx}
\hrefCMSnoop {} {P.~Nason, ``{A new method for combining NLO QCD with shower
  Monte Carlo algorithms}'',} \textit{ JHEP} \textbf{ 11} (2004) 040,
  \href{http://dx.doi.org/10.1088/1126-6708/2004/11/040}{\doi{10.1088/1126-6708/2004/11/040}},
\href{http://www.arXiv.org/abs/hep-ph/0409146}{\texttt{ arXiv:hep-ph/0409146}}.

\bibitem{Frixione:2007vw}
\hrefCMSnoop {} {S.~Frixione, P.~Nason, and C.~Oleari, ``{Matching NLO QCD
  computations with Parton Shower simulations: the POWHEG method}'',} \textit{
  JHEP} \textbf{ 11} (2007) 070,
  \href{http://dx.doi.org/10.1088/1126-6708/2007/11/070}{\doi{10.1088/1126-6708/2007/11/070}},
\href{http://www.arXiv.org/abs/0709.2092}{\texttt{ arXiv:0709.2092}}.

\bibitem{Madgraph5}
J.~Alwall\hrefCMSnoop {} { {et~al.}, ``{MadGraph 5}: going beyond'',} \textit{
  JHEP} \textbf{ 06} (2011) 128,
  \href{http://dx.doi.org/10.1007/JHEP06(2011)128}{\doi{10.1007/JHEP06(2011)128}},
\href{http://www.arXiv.org/abs/1106.0522}{\texttt{ arXiv:1106.0522}}.

\bibitem{Binoth:2008pr}
\hrefCMSnoop {} {T.~Binoth, N.~Kauer, and P.~Mertsch, ``{Gluon-induced QCD
  corrections to $\Pp\Pp \to \cPZ\cPZ \to \ell\bar{\ell}\ell'\bar{\ell'}$}'',}
  (2008).
\href{http://www.arXiv.org/abs/0807.0024}{\texttt{ arXiv:0807.0024}}.

\bibitem{CTEQ66}
H.-L. Lai\hrefCMSnoop {} { {et~al.}, ``Uncertainty induced by {QCD} coupling in
  the {CTEQ} global analysis of parton distributions'',} \textit{ Phys. Rev. D}
  \textbf{ 82} (2010) 054021,
  \href{http://dx.doi.org/10.1103/PhysRevD.82.054021}{\doi{10.1103/PhysRevD.82.054021}},
  \href{http://www.arXiv.org/abs/1004.4624}{\texttt{ arXiv:1004.4624}}.

\bibitem{ct10}
H.-L. Lai\hrefCMSnoop {} { {et~al.}, ``New parton distributions for collider
  physics'',} \textit{ Phys. Rev. D} \textbf{ 82} (2010) 074024,
  \href{http://dx.doi.org/10.1103/PhysRevD.82.074024}{\doi{10.1103/PhysRevD.82.074024}},
  \href{http://www.arXiv.org/abs/1007.2241}{\texttt{ arXiv:1007.2241}}.

\bibitem{Sherpa}
T.~Gleisberg\hrefCMSnoop {} { {et~al.}, ``{Event generation with \SHERPA
  1.1}'',} \textit{ JHEP} \textbf{ 02} (2009) 007,
  \href{http://dx.doi.org/10.1088/1126-6708/2009/02/007}{\doi{10.1088/1126-6708/2009/02/007}},
\href{http://www.arXiv.org/abs/0811.4622}{\texttt{ arXiv:0811.4622}}.

\bibitem{Jadach:1993hs}
S.~Jadach\hrefCMSnoop {} { {et~al.}, ``{The tau decay library TAUOLA: Version
  2.4}'',} \textit{ Comput. Phys. Commun.} \textbf{ 76} (1993) 361,
\href{http://dx.doi.org/10.1016/0010-4655(93)90061-G}{\doi{10.1016/0010-4655(93)90061-G}}.

\bibitem{GEANT}
\hrefCMSnoop {} {{ GEANT4} Collaboration, ``{GEANT4}--a simulation toolkit'',}
  \textit{ Nucl. Instrum. Meth. A} \textbf{ 506} (2003) 250,
\href{http://dx.doi.org/10.1016/S0168-9002(03)01368-8}{\doi{10.1016/S0168-9002(03)01368-8}}.

\bibitem{CMSExperiment}
\hrefCMSnoop {} {{ CMS} Collaboration, ``The {CMS} experiment at the {CERN}
  {LHC}'',} \textit{ JINST} \textbf{ 03} (2008) S08004,
\href{http://dx.doi.org/10.1088/1748-0221/3/08/S08004}{\doi{10.1088/1748-0221/3/08/S08004}}.

\bibitem{tracker}
\hrefCMSnoop {} {{ CMS} Collaboration, ``{CMS Tracking Performance Results from
  early LHC Operation}'',} \textit{ Eur. Phys. J. C} \textbf{ 70} (2010) 1165,
  \href{http://dx.doi.org/10.1140/epjc/s10052-010-1491-3}{\doi{10.1140/epjc/s10052-010-1491-3}},
\href{http://www.arXiv.org/abs/1007.1988}{\texttt{ arXiv:1007.1988}}.

\bibitem{Baffioni:2006cd}
S.~Baffioni\hrefCMSnoop {} { {et~al.}, ``{Electron reconstruction in CMS}'',}
  \textit{ Eur. Phys. J. C} \textbf{ 49} (2007) 1099,
\href{http://dx.doi.org/10.1140/epjc/s10052-006-0175-5}{\doi{10.1140/epjc/s10052-006-0175-5}}.

\bibitem{CMS-PAS-EGM-10-004}
\href {http://cdsweb.cern.ch/record/1299116} {{ CMS} Collaboration, ``Electron
  Reconstruction and Identification at $\sqrt{s} = 7$ {TeV}'',} CMS Physics
  Analysis Summary CMS-PAS-EGM-10-004, (2010).

\bibitem{CMS-PAS-MUO-10-002}
\hrefCMSnoop {} {{ CMS} Collaboration, ``{Performance of CMS muon
  reconstruction in pp collision events at $\sqrt{s}=7\TeV$}'',} \textit{
  JINST} \textbf{ 7} (2012) P10002,
  \href{http://dx.doi.org/10.1088/1748-0221/7/10/P10002}{\doi{10.1088/1748-0221/7/10/P10002}},
\href{http://www.arXiv.org/abs/1206.4071}{\texttt{ arXiv:1206.4071}}.

\bibitem{CMS:2011aa}
\hrefCMSnoop {} {{ CMS} Collaboration, ``{Measurement of the Inclusive $\PW$
  and $\PZ$ Production Cross Sections in pp Collisions at $\sqrt{s}=7\TeV$}'',}
  \textit{ JHEP} \textbf{ 10} (2011) 132,
  \href{http://dx.doi.org/10.1007/JHEP10(2011)132}{\doi{10.1007/JHEP10(2011)132}},
\href{http://www.arXiv.org/abs/1107.4789}{\texttt{ arXiv:1107.4789}}.

\bibitem{CMS-PAS-PFT-09-001}
\href {http://cdsweb.cern.ch/record/1194487} {{ CMS} Collaboration,
  ``Particle--Flow Event Reconstruction in {CMS} and Performance for Jets,
  Taus, and {\MET}'',} CMS Physics Analysis Summary CMS-PAS-PFT-09-001, (2009).

\bibitem{Chatrchyan:2011xq}
\hrefCMSnoop {} {{ CMS} Collaboration, ``{Performance of tau-lepton
  reconstruction and identification in CMS}'',} \textit{ JINST} \textbf{ 7}
  (2012) P01001,
  \href{http://dx.doi.org/10.1088/1748-0221/7/01/P01001}{\doi{10.1088/1748-0221/7/01/P01001}},
\href{http://www.arXiv.org/abs/1109.6034}{\texttt{ arXiv:1109.6034}}.

\bibitem{rhocorrection}
\hrefCMSnoop {} {M.~Cacciari and G.~Salam, ``Pileup substruction using jets'',}
  \textit{ Phys. Lett. B} \textbf{ 659} (2008) 119,
  \href{http://dx.doi.org/10.1016/j.physletb.2007.09.077}{\doi{10.1016/j.physletb.2007.09.077}},
\href{http://www.arXiv.org/abs/0707.1378}{\texttt{ arXiv:0707.1378}}.

\bibitem{CMS-EWK-TAU}
\hrefCMSnoop {} {{ CMS} Collaboration, ``Measurement of the Inclusive $Z$ Cross
  Section via Decays to Tau Pairs in $\Pp\Pp$ Collisions at
  $\sqrt{s}=7\TeV$'',} \textit{ JHEP} \textbf{ 08} (2011) 117,
  \href{http://dx.doi.org/10.1007/JHEP08(2011)117}{\doi{10.1007/JHEP08(2011)117}},
  \href{http://www.arXiv.org/abs/1104.1617}{\texttt{ arXiv:1104.1617}}.

\bibitem{MCFM}
\hrefCMSnoop {} {J.~M. Campbell and R.~K. Ellis, ``{MCFM for the Tevatron and
  the LHC}'',} \textit{ Nucl. Phys. Proc. Suppl.} \textbf{ 205} (2010) 10,
  \href{http://dx.doi.org/10.1016/j.nuclphysbps.2010.08.011}{\doi{10.1016/j.nuclphysbps.2010.08.011}},
\href{http://www.arXiv.org/abs/1007.3492}{\texttt{ arXiv:1007.3492}}.

\bibitem{Botje:2011sn}
M.~Botje\hrefCMSnoop {} { {et~al.}, ``{The PDF4LHC Working Group Interim
  Recommendations}'',} (2011).
\href{http://www.arXiv.org/abs/1101.0538}{\texttt{ arXiv:1101.0538}}.

\bibitem{Martin:2009iq}
A.~D. Martin\hrefCMSnoop {} { {et~al.}, ``{Parton distributions for the
  LHC}'',} \textit{ Eur. Phys. J. C} \textbf{ 63} (2009) 189,
  \href{http://dx.doi.org/10.1140/epjc/s10052-009-1072-5}{\doi{10.1140/epjc/s10052-009-1072-5}},
  \href{http://www.arXiv.org/abs/0901.0002}{\texttt{ arXiv:0901.0002}}.

\bibitem{nnpdf}
D.~R. Ball\hrefCMSnoop {} { {et~al.}, ``Impact of Heavy Quark Masses on Parton
  Distributions and LHC Phenomenology'',} \textit{ Nucl. Phys. B} \textbf{ 849}
  (2011) 296,
  \href{http://dx.doi.org/10.1016/j.nuclphysb.2011.03.021}{\doi{10.1016/j.nuclphysb.2011.03.021}},
  \href{http://www.arXiv.org/abs/1101.1300}{\texttt{ arXiv:1101.1300}}.

\bibitem{lumiPAS12}
\href {http://cdsweb.cern.ch/record/1434360} {{ CMS} Collaboration, ``Absolute
  Calibration of the Luminosity Measurement at {CMS}: {W}inter 2012 Update'',}
  CMS Physics Analysis Summary CMS-PAS-SMP-12-008, (2012).

\bibitem{CLS1}
\hrefCMSnoop {} {T.~Junk, ``{Confidence level computation for combining
  searches with small statistics}'',} \textit{ Nucl. Instrum. Meth. A} \textbf{
  434} (1999) 435,
  \href{http://dx.doi.org/10.1016/S0168-9002(99)00498-2}{\doi{10.1016/S0168-9002(99)00498-2}},
  \href{http://www.arXiv.org/abs/hep-ex/9902006}{\texttt{
  arXiv:hep-ex/9902006}}.

\bibitem{CLS2}
\hrefCMSnoop {} {A.~L. Read, ``{Presentation of search results: the $CL_s$
  technique}'',} \textit{ J. Phys. G} \textbf{ 28} (2002) 2693,
\href{http://dx.doi.org/10.1088/0954-3899/28/10/313}{\doi{10.1088/0954-3899/28/10/313}}.

\bibitem{ATL-PHYS-PUB-2011-011}
\href {https://cdsweb.cern.ch/record/1379837} {{ATLAS and CMS Collaborations},
  ``Procedure for the LHC Higgs boson search combination in summer 2011'',}
  ATL-PHYS-PUB-2011-011, {CMS NOTE-2011/005}, (2011).

\bibitem{Zralek:1989ty}
\href {http://www.actaphys.uj.edu.pl/vol20/pdf/v20p0739.pdf} {M.~Zralek and
  P.~Khristova, ``{Composite Z boson in the $\Pe^+\Pe^- \to \cPZ\cPZ$
  process}'',} \textit{ Acta Phys. Polon. B} \textbf{ 20} (1989)
739.

\end{thebibliography}\endgroup
